\documentclass[a4paper,aps,prd,10pt,preprintnumbers,showpacs,twocolumn,superscriptaddress,nofootinbib,amsmath,amssymb,floatfix]{revtex4-1}
\usepackage{graphicx}
\usepackage{cmap}
\usepackage[utf8]{inputenc}
\usepackage[T1]{fontenc}
\usepackage[colorlinks=true,linktocpage=true,linkcolor=blue,citecolor=blue,allcolors=blue]{hyperref}
\usepackage{color}

\begin{document}
\title{Quasinormal Modes and Gray-Body Factors for Gravitational Perturbations in Asymptotically Safe Gravity}
\author{B. C. Lütfüoğlu}
\email{bekir.lutfuoglu@uhk.cz}
\affiliation{Department of Physics, Faculty of Science, University of Hradec Kralove, \\
Rokitanskeho 62/26, Hradec Kralove, 500 03, Czech Republic. }
\begin{abstract}
A quantum-corrected black hole model arising from gravitational collapse in the framework of Asymptotically Safe Gravity was recently proposed in [A. Bonanno, D. Malafarina, A. Panassiti, Phys. Rev. Lett. 132, 031401 (2024)]. Quantum correction becomes considerable for  Planck-scale black holes and strongly deviates from the Schwarzschild solution near the event horizon, quickly merging with the Schwarzschild metric in the far region. While quasinormal modes and gray-body factors have been analyzed for test fields in this background, no such analysis has yet been performed for gravitational perturbations. In this work, we study axial gravitational perturbations of these black holes by modeling the effective quantum corrections through an anisotropic fluid energy-momentum tensor. We compute both quasinormal modes and gray-body factors, and show that quantum corrections enhance the quality factor of the oscillations, thereby making the quantum-corrected black hole a more efficient gravitational wave emitter. At asymptotically late times, the power-law decay is indistinguishable from Price's tails, which behave as \( \sim t^{-(2\ell + 3)} \), where \( \ell \) is the multipole number. We also demonstrate that quantum corrections lead to a suppression of the gray-body factors and examine the validity of the correspondence between gray-body factors and quasinormal modes.

\end{abstract}
\maketitle
\section{Introduction}

The detection of gravitational waves from binary black hole mergers by the LIGO and Virgo collaborations~\cite{Abbott:2016blz,Abbott:2016nmj,LIGOScientific:2017vwq,LIGOScientific:2020zkf} has ushered in a new era in observational astrophysics, providing unprecedented insights into the strong-field regime of gravity. The ringdown phase of these events, characterized by quasinormal modes (QNMs), offers a unique opportunity to probe the near-horizon geometry of black holes and test the predictions of general relativity (see  \cite{Kokkotas:1999bd,Berti:2009kk,Konoplya:2011qq,Bolokhov:2025uxz} for reviews).

QNMs are intrinsic oscillations of perturbed black holes, with complex frequencies whose real parts correspond to oscillation frequencies and imaginary parts to damping rates. These modes are determined solely by the black hole's parameters and the underlying gravitational theory, making them powerful tools for exploring modifications to general relativity and the nature of spacetime singularities.

Regular black hole solutions, which avoid central singularities by incorporating quantum gravitational effects or alternative matter sources, have garnered significant attention~\cite{Bardeen:1968,Hayward:2006}. In particular, models inspired by Asymptotically Safe Gravity \cite{Niedermaier:2006wt} propose modifications to the Schwarzschild solution that remain regular at the core~\cite{Bonanno:2000ep,Held:2019xde,Platania:2019kyx}. Recent studies have investigated the QNMs of various fields in these backgrounds~\cite{Saleh:2016pke,Saleh:2014uca,Konoplya:2023aph,Konoplya:2022hll,Zinhailo:2023xdz}, revealing distinctive features in the fundamental mode and overtone spectra.

A recent work~\cite{Bonanno:2023rzk} derived a fully regular black hole solution by modeling gravitational collapse in the framework of Asymptotically Safe Gravity. Unlike many earlier approaches that introduce regularity through ad hoc modifications of the static metric or the Misner–Sharp mass, this model arises from an effective Lagrangian with a running gravitational coupling, guided by the Reuter fixed point, and ensures geodesic completeness without violating energy conservation. The resulting black hole is sourced by a dust interior and matched consistently to a static exterior through Israel junction conditions, yielding a novel, singularity-free geometry with quantum-improved dynamics.
According to the renormalization group approach and introduction of the cut-off parameter, the quantum correction becomes significant when the mass of the black hole $M$ is close to the Planck mass $M_{Pl}$. For large astrophysical black holes the dimensionless quantum parameter $\xi/M^2$ is small, so that the deviation from the classical geometry is negligible.

QNMs of test scalar, electromagnetic, and Dirac fields in the background of such black holes have been recently analyzed in \cite{Stashko:2024wuq}. However, to fully understand the stability and observational signatures of these regular black holes, it is essential to analyze the QNMs arising from gravitational perturbations. Gravitational QNMs are directly related to the emission of gravitational waves and thus have immediate relevance for current and future observations. Moreover, they can provide deeper insights into the dynamical response of the spacetime itself, beyond the behavior of test fields.

In this work, we focus on the gravitational perturbations of regular black holes within the framework of asymptotically safe gravity. By deriving and solving the master equations governing these perturbations, we aim to compute the corresponding QNM spectra.

The paper is organized as follows. In Sec. \ref{sec:wavelike}, we briefly review the quantum-corrected black hole solution arising in asymptotically safe gravity and derive the master equation for axial gravitational perturbations, modeling the quantum corrections as an effective anisotropic fluid. In Sec. \ref{sec:WKB}, we describe the computational methods employed for extracting QNMs and gray-body factors, including the higher-order JWKB approach with Padé approximants and the time-domain integration method. The results for quasinormal frequencies and their accuracy are discussed in Sec. \ref{sec:QNMs}, while Sec. \ref{sec:GBF} is devoted to the analysis of gray-body factors and their relation to QNMs. Finally, in Sec.  \ref{sec:Conc} we summarize our findings and comment on possible directions for future work.

\section{Metric, Perturbation equations and effective potentials}\label{sec:wavelike}

The metric of the Bonanno-Malafarina-Panassiti (BMP) black hole is given by the following line element \cite{Bonanno:2023rzk},
\begin{equation}\label{metric}
  ds^2=-f(r)dt^2+\frac{dr^2}{f(r)}+r^2(d\theta^2+\sin^2\theta d\phi^2),
\end{equation}
where
$$
\begin{array}{rcl}
f(r)&=&\displaystyle 1-\frac{r^2 \log \left(\frac{6 \xi  M}{r^3}+1\right)}{3 \xi }.
\end{array}
$$
Here $\xi $ is the coupling, which has dimensionality of squared mass and $M$ is the Arnowitt–Deser–Misner mass. We will further measure all dimensional quantities in units of the mass $M=1$. Various optical observational phenomena for this black hole metric have been considered in \cite{Urmanov:2025nou}.

In the renormalization-group (RG) improved approach to asymptotically safe gravity, the parameter $\xi$ plays the role of a matching constant that links the RG scale $k$ to the local energy density $\varepsilon$ of the collapsing matter. This identification allows one to express the running Newton coupling and effective cosmological term directly as functions of $\varepsilon$. After setting $8\pi G_N=1$, the resulting density-dependent couplings take the compact form \cite{Bonanno:2023rzk}:
\begin{widetext}
\begin{equation}
G(\varepsilon) \sim \frac{1}{1+\xi\,\varepsilon}, \qquad
\chi(\varepsilon) = \frac{\ln(1+\xi\,\varepsilon)}{\xi\,\varepsilon}, \qquad
\Lambda(\varepsilon) = \frac{\ln(1+\xi\,\varepsilon)}{\xi} - \frac{\varepsilon}{1+\xi\,\varepsilon}.
\end{equation}
\end{widetext}
Here $\chi(\varepsilon)$ modifies the matter sector, while $\Lambda(\varepsilon)$ behaves as an induced, density-dependent cosmological term.  Physically, $\xi$ determines the density threshold at which quantum corrections become significant: for $\xi \rightarrow 0$ the classical general-relativistic limit is recovered, whereas finite $\xi$ introduces nontrivial quantum-gravity effects already at lower densities.

The exterior Misner--Sharp mass for the model of~\cite{Bonanno:2023rzk} is given by
\begin{equation}
M(R) = \frac{R^{3}}{6\,\xi}\,\ln\!\left(1+\frac{6\,M_{0}\,\xi}{R^{3}}\right),
\end{equation}
so that the metric function can be written as
\begin{equation}
f(R) = 1 - \frac{2M(R)}{R}.
\end{equation}
A dimensional analysis in geometrical units ($G=c=1$) shows that $R$ and $M_{0}$ both have dimensions of length.
Since the argument of the logarithm must be dimensionless, one finds
\begin{equation}
\frac{M_{0}\,\xi}{R^{3}} \sim 1 \quad \Rightarrow \quad [\xi] = L^{2},
\end{equation}
so that $\xi$ has the dimensions of length$^{2}$ (or, equivalently, mass$^{2}$).

In line with the approach developed in~\cite{Bouhmadi-Lopez:2020oia,Konoplya:2024lch}, we study axial gravitational perturbations around a spherically symmetric black hole modified by quantum effects. Since the background geometry is derived from an effective framework rather than being a direct solution of the Einstein field equations, quantum corrections are modeled through an effective stress-energy tensor describing an anisotropic fluid. As argued in~\cite{Ashtekar:2018cay,Bouhmadi-Lopez:2020oia,Konoplya:2024lch}, it is reasonable —particularly for geometries that closely resemble Schwarzschild spacetime — to omit perturbations along the anisotropic direction in the linearized treatment.

The metric perturbation in the Regge–Wheeler gauge is given by:
\begin{equation}
h^{\text{axial}}_{\mu\nu} =
\begin{pmatrix}
0 & 0 & 0 & h_0(t,r) \\
0 & 0 & 0 & h_1(t,r) \\
0 & 0 & 0 & 0 \\
h_0(t,r) & h_1(t,r) & 0 & 0
\end{pmatrix}
\sin\theta \, \partial_\theta P_\ell(\cos\theta),
\end{equation}
where \( h_0(t, r) \) and \( h_1(t, r) \) are the metric perturbation functions and \( P_\ell(\cos\theta) \) is the Legendre polynomial.

The quantum corrections to gravity are modeled as an anisotropic fluid with stress-energy tensor:
\begin{equation}
T^{\mu\nu} = (\rho + p_t) u^\mu u^\nu + p_t g^{\mu\nu} + (p_r - p_t) s^\mu s^\nu,
\end{equation}
where \( \rho \), \( p_r \), and \( p_t \) are the energy density, radial pressure, and tangential pressure, respectively. The four-velocity \( u^\mu \) and radial unit vector \( s^\mu \) are defined as:
\begin{equation}
u^\mu = (\sqrt{f(r)}, 0, 0, 0), \qquad s^\mu = \left(0, \frac{1}{\sqrt{f(r)}}, 0, 0\right),
\end{equation}
satisfying the orthonormality conditions:
\begin{equation}
u^\mu u_\mu = -1, \quad s^\mu s_\mu = 1, \quad u^\mu s_\mu = 0.
\end{equation}

Since \( \rho \), \( p_r \), and \( p_t \) are scalar quantities, their axial perturbations vanish. Perturbations of \( u^\mu \) and \( s^\mu \) have non-zero components:
\begin{align}
\delta u^\phi &= -i \omega U(r) e^{-i\omega t} \sin\theta \, \partial_\theta P_\ell(\cos\theta), \\
\delta s^\phi &= -S(r) e^{-i\omega t} \sin\theta \, \partial_\theta P_\ell(\cos\theta).
\end{align}

Assuming \( \delta s^\mu = 0 \), the conservation equation \( \nabla_\mu T^{\mu r} = 0 \) implies \( \delta u^\phi = 0 \). Substituting into the linearized Einstein equations leads to:
\begin{widetext}
\begin{align}
h_1(r) \left[r^2 \omega^2 - (\ell - 1)(\ell + 2) f(r)\right] - i r^2 \omega h_0'(r) + 2 i r \omega h_0(r) &= 0, \\
f(r) \left[\frac{f'(r)}{f(r)} h_1(r) + 2 h_1'(r)\right] + \frac{2 i \omega}{f(r)} h_0(r) &= 0.
\end{align}
\end{widetext}

We introduce the tortoise coordinate \( r_* \) and a new perturbation variable \( \Psi(r) \) via:
\begin{equation}
h_1(r) = \frac{r}{f(r)} \Psi(r), \qquad \frac{dr_*}{dr} = \frac{1}{f(r)}.
\end{equation}

This leads to the master wave equation for axial perturbations:
\begin{equation}\label{masterequation}
\frac{d^2 \Psi}{dr_*^2} + \left[\omega^2 - V(r)\right] \Psi = 0,
\end{equation}
with the effective potential given by:
\begin{equation}
V(r) = f(r) \left[\frac{2}{r^2} - \frac{f'(r)}{2 r f(r)} + \frac{(\ell + 2)(\ell - 1)}{r^2} \right].
\end{equation}

This expression reduces to the classical Regge–Wheeler potential in the Schwarzschild limit but captures modifications from quantum corrections through the modified form of \( f(r) \). At the same time, following \cite{Chakraborty:2024gcr}, one finds that if perturbations along the anisotropy direction are neglected, the corresponding perturbations of the energy–momentum tensor vanish. The complete effective potential derived in \cite{Chakraborty:2024gcr} also involves the pressure, thereby implying the existence of a fixed equation of state. Such an assumption, however, is difficult to justify when dealing with a phenomenologically constructed black-hole metric. 

\begin{figure}
\resizebox{\linewidth}{!}{\includegraphics{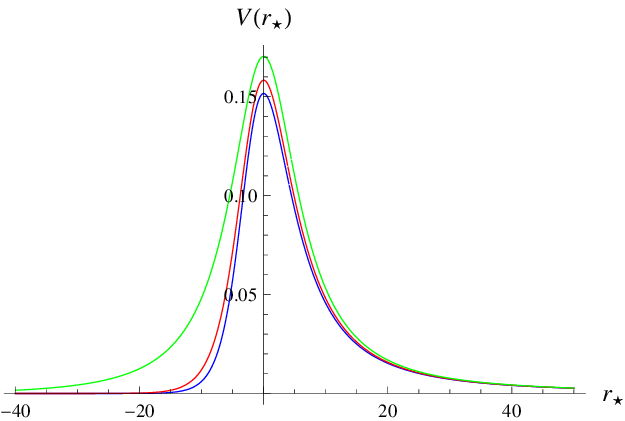}}
\caption{Effective potential as a function of the tortoise coordinate $r^{*}$ for $\ell=2$, $\xi=0.01$ (blue), $\xi=0.2$ (red), $\xi=0.45$ (green).}\label{fig:gravpot1}
\end{figure}

\begin{figure}
\resizebox{\linewidth}{!}{\includegraphics{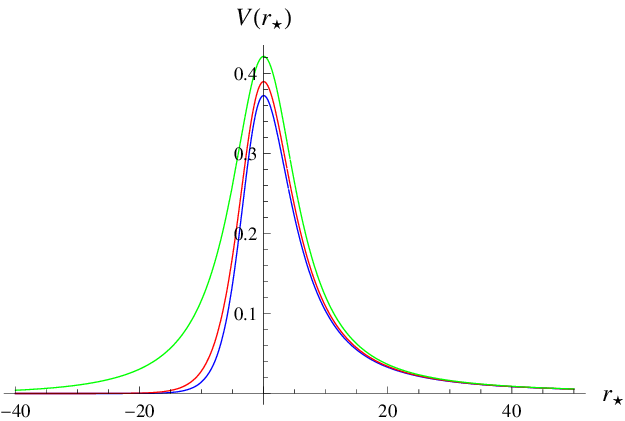}}
\caption{Effective potential as a function of the tortoise coordinate $r^{*}$ for $\ell=3$, $\xi=0.01$ (blue), $\xi=0.2$ (red), $\xi=0.45$ (green).}\label{fig:gravpot2}
\end{figure}

The examples of effective potentials for $\ell=2$ and $\ell=3$ perturbations for various values of the quantum parameter $\xi$ are shown in Figs.  \ref{fig:gravpot1} and \ref{fig:gravpot2}. The effective potentials are positive definite everywhere outside the event horizon, so that the perturbations are stable and there must be no growing modes in the spectrum.

\section{JWKB approach and time-domain integration}\label{sec:WKB}

The QNMs correspond to solutions of Eq.~\eqref{masterequation} satisfying purely ingoing and outgoing boundary conditions:
\begin{equation}
\Psi(r_*) \sim
\begin{cases}
e^{-i\omega r_*}, & r_* \to -\infty \quad (\text{near horizon}), \\
e^{+i\omega r_*}, & r_* \to +\infty \quad (\text{spatial infinity}).
\end{cases}
\label{qnms_bc}
\end{equation}
These conditions describe waves that fall into the black hole at the horizon and radiate outward at spatial infinity, with no incoming radiation.

For gray-body factors, one considers scattering problems where an incoming wave from infinity is partially transmitted into the black hole and partially reflected back:
\begin{equation}
\Psi(r_*) \sim
\begin{cases}
T e^{-i\omega r_*}, & r_* \to -\infty, \\
e^{-i\omega r_*} + R e^{+i\omega r_*}, & r_* \to +\infty,
\end{cases}
\label{greybody_bc}
\end{equation}
where \( T \) and \( R \) are the transmission and reflection amplitudes. The \textit{gray-body factor} is then defined by
\begin{equation}
\Gamma(\omega) = |T|^2.
\end{equation}
To compute the QNMs or transmission/reflection coefficients, we employ the JWKB method extended to higher orders and refined by Padé approximants. The JWKB quantization condition at \( N \)-th order is given by~\cite{Iyer:1986np,Konoplya:2003ii,Matyjasek:2017psv}:
\begin{equation}
\frac{i Q_0}{\sqrt{2 Q_0''}} - \sum_{j=2}^{N} \Lambda_j = n + \frac{1}{2}, \quad n = 0,1,2,\dots,
\label{wkb_formula}
\end{equation}
where \( Q(r) = \omega^2 - V(r) \), and \( Q_0 \), \( Q_0'' \) denote the value and second derivative of \( Q(r) \) at the peak of the potential. The \( \Lambda_j \) are higher-order correction terms, expressed in terms of derivatives of \( V(r) \) up to order \( 2N \).  The accuracy of the JWKB expansion can be significantly improved using Padé approximants \cite{Matyjasek:2017psv}, denoted as \( P^m_n \), applied to the JWKB series. For instance, the sixth-order JWKB series can be converted into a Padé approximant \( P^6_6 \) to obtain more accurate QNM frequencies. The choice of the Padé order \( m \) is usually fixed by optimizing the agreement with exact results in known limits (e.g., Schwarzschild). The JWKB method is frequently used at various orders in order to find QNMs and gray-body factors  \cite{Kodama:2009bf,Konoplya:2005sy,Konoplya:2001ji,Konoplya:2006ar,Kokkotas:2010zd,Ishihara:2008re,Zhao:2022gxl,Bolokhov:2023ruj,Momennia:2022tug,Barrau:2019swg,DuttaRoy:2022ytr}.

In the context of gray-body factors, the JWKB approach also provides approximate expressions for the reflection and transmission coefficients when the potential has the standard barrier shape. At leading order (third order or higher is more accurate), the reflection coefficient can be written as:
\begin{equation}
|R|^2 = \left( 1 + e^{-2\pi K} \right)^{-1},
\end{equation}
where \( K \) depends on the shape of the potential barrier and includes JWKB correction terms. Then, the gray-body factor is
\begin{equation}
\Gamma(\omega) = 1 - |R|^2 = \left( 1 + e^{2\pi K} \right)^{-1}.
\end{equation}
This approach yields accurate results when the potential is well-behaved and single-peaked. The JWKB-Padé method is particularly effective for low overtone QNMs and frequencies near the peak of the barrier, which are most relevant for observable gravitational wave signals.

The method works best when:
\begin{itemize}
  \item The effective potential is positive definite and has a single peak.
  \item The overtone number \( n \) is less than or comparable to the multipole number \( \ell \).
\end{itemize}

In cases where the potential deviates significantly from the standard shape or possesses multiple extrema (e.g., massive fields or exotic matter distributions), the JWKB approximation may become unreliable, and numerical integration or continued fraction methods are more appropriate.

The time-domain integration method, developed by Gundlach, Price, and Pullin~\cite{Gundlach:1993tp}, is a numerical technique for evolving wave-like perturbation equations in black hole spacetimes. The basic idea is to solve the master equation
\begin{equation}
\left( \frac{\partial^2}{\partial t^2} - \frac{\partial^2}{\partial r_*^2} + V(r) \right) \Psi(t, r) = 0,
\end{equation}
using light-cone coordinates \( u = t - r_* \), \( v = t + r_* \). The wave equation is discretized on a uniform grid in the \( (u, v) \) plane using finite difference schemes. A common discretization is the second-order accurate scheme:
\begin{align}
\Psi(N) &= \Psi(W) + \Psi(E) - \Psi(S) \notag \\
&\quad - \frac{\Delta^2}{8} \left[ V(W)\Psi(W) + V(E)\Psi(E) \right],
\end{align}
where \( N, S, E, W \) denote grid points to the North, South, East, and West, and \( \Delta \) is the grid spacing. Initial data are specified on two null surfaces (e.g., a Gaussian pulse), and the evolution proceeds step-by-step to generate the full time-domain signal \( \Psi(t, r) \). Time-domain integration method has been applied in numerous publications not only for finding QNMs, but also for testing stability of perturbations and analysis of asymptotic tails \cite{Konoplya:2006gq,Varghese:2011ku,Bolokhov:2024ixe,Hamil:2025vey,Lutfuoglu:2025hjy,Qian:2022kaq,Konoplya:2020jgt,Dubinsky:2024hmn,Dubinsky:2024aeu,Churilova:2021tgn,Dubinsky:2024gwo,Churilova:2019qph,Konoplya:2013sba,Lutfuoglu:2025hwh,Skvortsova:2024wly}

To extract QNMs from the resulting waveform, one typically employs the Prony method, which fits the late-time signal to a superposition of damped sinusoids:
\begin{equation}\nonumber
\Psi(t) \approx \sum_{n=0}^{N-1} A_n e^{-i \omega_n t}, \quad \omega_n = \text{Re}(\omega_n) - i\, \text{Im}(\omega_n),
\end{equation}
where \( A_n \) are amplitudes and \( \omega_n \) are complex frequencies. The Prony algorithm reconstructs \( \omega_n \) and \( A_n \) by solving a linear system derived from the sampled data. This method allows for the precise determination of the fundamental modes and testing the (in)stability.

\begin{figure*}
\resizebox{\linewidth}{!}{\includegraphics{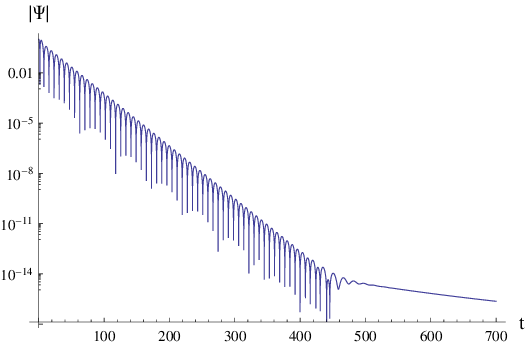}~~\includegraphics{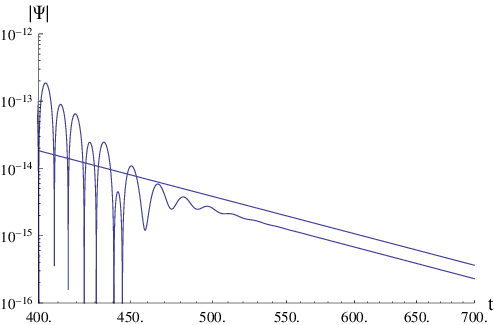}}
\caption{Left panel: Semi-logarithmic time-domain profile for $\ell=2$, $\xi=0.45$, $M=1$. The fundamental mode given by the Prony method is $\omega = 0.40033 - 0.07234 i$, which is in excellent concordance with the JWKB result $\omega =  0.400330 - 0.072338 i $. Right panel: The asymptotic tail (logarithmic scale) for the same values of the parameters together with the line $|\Psi| = A \cdot t^{-7}$, where $A$ is a constant.}\label{fig:TD1}
\end{figure*}

\begin{figure*}
\resizebox{\linewidth}{!}{\includegraphics{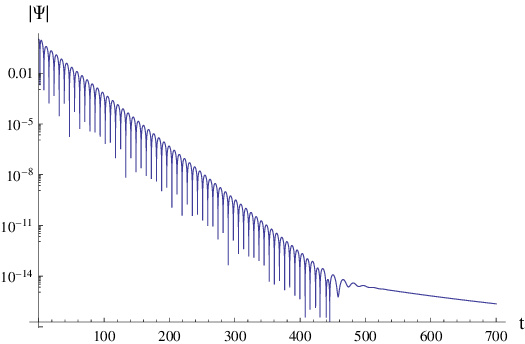}~~\includegraphics{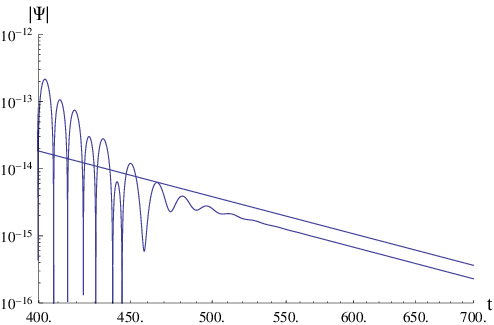}}
\caption{Left panel: Semi-logarithmic time-domain profile for $\ell=2$, $\xi=0.455$, $M=1$.
The fundamental mode given by the Prony method is $\omega = 0.400669 - 0.0720145 i$, which is in excellent concordance with the JWKB result $\omega =  0.400668 - 0.072007 i$. Right panel: The asymptotic tail (logarithmic scale) for the same values of the parameters together with the line $|\Psi| = A \cdot t^{-7}$, where $A$ is a constant.}\label{fig:TD2}
\end{figure*}

\begin{table}
\begin{tabular}{c c c c}
\hline
\hline
$\xi$ & JWKB6 $m=3$ & JWKB7 $m=4$ & difference  \\
\hline
$0.01$ & $0.374134-0.088714 i$ & $0.374148-0.088755 i$ & $0.0111\%$\\
$0.05$ & $0.376176-0.087835 i$ & $0.376150-0.087843 i$ & $0.0070\%$\\
$0.1$ & $0.378773-0.086611 i$ & $0.378743-0.086610 i$ & $0.00780\%$\\
$0.15$ & $0.381496-0.085161 i$ & $0.381448-0.085248 i$ & $0.0253\%$\\
$0.2$ & $0.384220-0.083700 i$ & $0.384272-0.083745 i$ & $0.0175\%$\\
$0.25$ & $0.387196-0.082045 i$ & $0.387229-0.082061 i$ & $0.00925\%$\\
$0.3$ & $0.390309-0.080147 i$ & $0.390317-0.080148 i$ & $0.00186\%$\\
$0.35$ & $0.393542-0.077965 i$ & $0.393554-0.077962 i$ & $0.00317\%$\\
$0.4$ & $0.396903-0.075348 i$ & $0.396919-0.075388 i$ & $0.0107\%$\\
$0.45$ & $0.400330-0.072338 i$ & $0.400324-0.072349 i$ & $0.00293\%$\\
$0.455$ & $0.400668-0.072007 i$ & $0.400663-0.072016 i$ & $0.00255\%$\\
\hline
\hline
\end{tabular}
\caption{QNMs of the $\ell=2$, $n=0$ gravitational field  for the BMP black hole $M=1$ calculated using the JWKB formula at different orders and Padé approximants.}\label{table1}
\end{table}

\begin{table}
\begin{tabular}{c c c c}
\hline
\hline
$\xi$ & JWKB6 $m=3$ & JWKB7 $m=4$ & difference  \\
\hline
$0.01$ & $0.347108-0.272753 i$ & $0.347514-0.273453 i$ & $0.183\%$\\
$0.05$ & $0.350568-0.270045 i$ & $0.350478-0.270196 i$ & $0.0397\%$\\
$0.1$ & $0.354610-0.265736 i$ & $0.354329-0.265952 i$ & $0.0801\%$\\
$0.15$ & $0.359097-0.260333 i$ & $0.358218-0.261335 i$ & $0.300\%$\\
$0.2$ & $0.359613-0.252869 i$ & $0.362117-0.256256 i$ & $0.958\%$\\
$0.25$ & $0.364221-0.250314 i$ & $0.365917-0.250553 i$ & $0.388\%$\\
$0.3$ & $0.368580-0.243977 i$ & $0.369480-0.244080 i$ & $0.205\%$\\
$0.35$ & $0.372119-0.236644 i$ & $0.372594-0.236718 i$ & $0.109\%$\\
$0.4$ & $0.374599-0.228410 i$ & $0.374939-0.228437 i$ & $0.0775\%$\\
$0.45$ & $0.375908-0.219551 i$ & $0.376004-0.219473 i$ & $0.0283\%$\\
$0.455$ & $0.375957-0.218619 i$ & $0.376022-0.218557 i$ & $0.0205\%$\\
\hline
\hline
\end{tabular}
\caption{QNMs of the $\ell=2$, $n=1$ gravitational field  for the BMP black hole $M=1$ calculated using the JWKB formula at different orders and Padé approximants.}\label{table2}
\end{table}

\begin{table}
\begin{tabular}{c c c c}
\hline
\hline
$\xi$ & JWKB6 $m=3$ & JWKB7 $m=4$ & difference  \\
\hline
$0.01$ & $0.600163-0.092490 i$ & $0.600162-0.092490 i$ & $0.00010\%$\\
$0.05$ & $0.603107-0.091601 i$ & $0.603105-0.091600 i$ & $0.00028\%$\\
$0.1$ & $0.606943-0.090392 i$ & $0.606943-0.090391 i$ & $0.00019\%$\\
$0.15$ & $0.610976-0.089058 i$ & $0.610976-0.089058 i$ & $0.00005\%$\\
$0.2$ & $0.615225-0.087574 i$ & $0.615226-0.087574 i$ & $0.00011\%$\\
$0.25$ & $0.619716-0.085906 i$ & $0.619716-0.085908 i$ & $0.00040\%$\\
$0.3$ & $0.624474-0.084011 i$ & $0.624477-0.084013 i$ & $0.00056\%$\\
$0.35$ & $0.629538-0.081826 i$ & $0.629539-0.081826 i$ & $0.00021\%$\\
$0.4$ & $0.634931-0.079258 i$ & $0.634931-0.079258 i$ & $0\%$\\
$0.45$ & $0.640668-0.076178 i$ & $0.640668-0.076178 i$ & $0\%$\\
$0.455$ & $0.641261-0.075836 i$ & $0.641261-0.075836 i$ & $0\%$\\
\hline
\hline
\end{tabular}
\caption{QNMs of the $\ell=3$, $n=0$ gravitational field  for the BMP black hole $M=1$ calculated using the JWKB formula at different orders and Padé approximants.}\label{table3}
\end{table}

\begin{table}
\begin{tabular}{c c c c}
\hline
\hline
$\xi$ & JWKB6 $m=3$ & JWKB7 $m=4$ & difference  \\
\hline
$0.01$ & $0.583508-0.280625 i$ & $0.583508-0.280625 i$ & $0\%$\\
$0.05$ & $0.587029-0.277809 i$ & $0.587026-0.277805 i$ & $0.00085\%$\\
$0.1$ & $0.591547-0.273977 i$ & $0.591546-0.273973 i$ & $0.00053\%$\\
$0.15$ & $0.596201-0.269742 i$ & $0.596201-0.269742 i$ & $0\%$\\
$0.2$ & $0.600984-0.265029 i$ & $0.600990-0.265039 i$ & $0.00184\%$\\
$0.25$ & $0.605867-0.259738 i$ & $0.605874-0.259759 i$ & $0.00328\%$\\
$0.3$ & $0.610797-0.253741 i$ & $0.610805-0.253771 i$ & $0.00463\%$\\
$0.35$ & $0.615664-0.246880 i$ & $0.615674-0.246913 i$ & $0.00523\%$\\
$0.4$ & $0.620268-0.238971 i$ & $0.620266-0.239003 i$ & $0.00484\%$\\
$0.45$ & $0.624205-0.229880 i$ & $0.624194-0.229911 i$ & $0.00491\%$\\
$0.455$ & $0.624541-0.228906 i$ & $0.624530-0.228937 i$ & $0.00490\%$\\
\hline
\hline
\end{tabular}
\caption{QNMs of the $\ell=3$, $n=1$ gravitational field  for the BMP black hole $M=1$ calculated using the JWKB formula at different orders and Padé approximants.}\label{table4}
\end{table}

\section{Quasinormal modes}\label{sec:QNMs}

In Tables~\ref{table1}, \ref{table2}, \ref{table3}, and \ref{table4}, we present the quasinormal frequencies of the fundamental mode and the first overtone for various values of the coupling constant \( \xi \), ranging from the nearly Schwarzschild limit \( \xi = 0.01 \) up to the near-extremal value \( \xi = 0.455 \) (in units where \( M = 1 \)). We observe that the real part of the frequency increases monotonically with \( \xi \), while the damping rate (i.e., the absolute value of the imaginary part) decreases monotonically. Consequently, the quality factor,
\begin{equation}
Q \sim \left| \frac{\text{Re}\,\omega}{\text{Im}\,\omega} \right|,
\end{equation}
grows as quantum corrections are introduced.

Although the JWKB method is asymptotic in nature, it has been shown in various works~\cite{Konoplya:2019hlu,Lutfuoglu:2025hwh,Malik:2025ava,Konoplya:2020jgt} that if the results at successive JWKB orders are close — forming a plateau of "local convergence" — then the method is considered stable, and the relative error can typically be estimated as the difference between successive orders. In our case, the relative difference does not exceed approximately \( 0.1\% \), and in most instances, it is even smaller by several orders of magnitude. This error estimate is clearly much smaller than the physical effect we aim to resolve: the deviations of the frequencies from their Schwarzschild values reach up to \( 20\% \) for the damping rate and around \(  7\% \) for the real part. Notably, the imaginary part of the frequency is much more sensitive to quantum corrections than the real part, indicating that damping is more significantly affected than the oscillation frequency.

In addition to comparing different JWKB orders, the accuracy of the JWKB method can also be assessed by cross-verifying with time-domain integration. This numerical approach enables a precise determination of the fundamental mode (\( n = 0 \)). As shown in Figs.~\ref{fig:TD1} and \ref{fig:TD2}, the difference between the frequencies obtained via the JWKB method and those extracted from time-domain integration is of the order of \( 10^{-6} \) or smaller. For example, for $\xi=0.455$, the fundamental mode given by the Prony method is $\omega = 0.641263 - 0.0758349 i$, which is in excellent concordance with the JWKB result $0.641261 - 0.075836 i$. This excellent agreement confirms the high precision of the JWKB results and validates their use for clearly identifying the effects introduced by quantum corrections. For all the data presented in the tables, except for the smallest considered value $\xi = 0.01$, the relative error is at least one order of magnitude smaller than the physical effect under study, namely the deviation from the Schwarzschild limit. In the Schwarzschild case, for $\ell=2$, the precise fundamental frequency is $\omega = 0.373672 - 0.0889625 i$, which is in very close agreement with the WKB result $\omega = 0.374134 - 0.088714 i$ for $\xi =0.01$. In the Schwarzschild limit $\xi=0$, the WKB method with Padé approximants reproduces the quasinormal frequencies with an accuracy of up to five decimal places \cite{Matyjasek:2017psv}.

In the eikonal limit, we can represent the position of the maximum of the effective potential as a series in terms of small $\ell^{-1}$ and $\xi$,
\begin{equation}
r_{max} =  \left(3 M-\frac{2 \xi }{3 M}-\frac{8 \xi ^2}{27
   M^3}\right)+O\left(\frac{1}{\kappa^2
   }, \xi^3 \right).
\end{equation}
Then, using this relation in the first-order JWKB formula, we find the analytic expressions for the frequencies in the regime of high multipole numbers:
\begin{align}
\omega_{n} = &\kappa \left( \frac{1}{3 \sqrt{3} M} + \frac{\xi}{27 \sqrt{3} M^3}
+ \frac{25 \xi^2}{1458 \sqrt{3} M^5}  \right) \notag \\
& -\frac{i K}{3 \sqrt{3} M} + \frac{2 i K \xi}{27 \sqrt{3} M^3}
+ \frac{46 i K \xi^2}{729 \sqrt{3} M^5}
+ O\left( \frac{1}{\kappa} , \xi^3 \right),
\end{align}
where
\begin{equation}
K=n+\frac{1}{2}, \quad \kappa =\ell +\frac{1}{2}.
\end{equation}

\begin{figure*}
\resizebox{\linewidth}{!}{\includegraphics{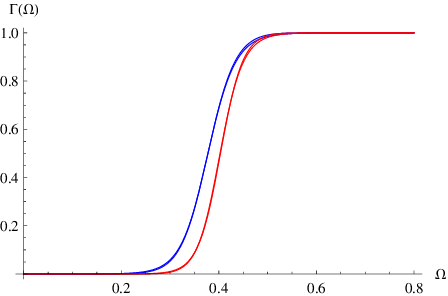}\includegraphics{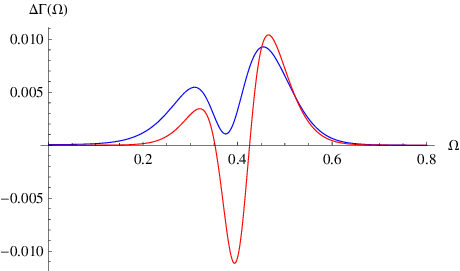}}
\caption{Left panel: Gray-body factors obtained by the 6th order JWKB approach and via the correspondence with QNMs $\ell=2$, $\xi=0.01$ (blue) and $\xi=0.45$ (red); $M=1$. Right panel: Difference between gray-body factors from both methods.}\label{fig:GBFL2}
\end{figure*}
\begin{figure*}
\resizebox{\linewidth}{!}{\includegraphics{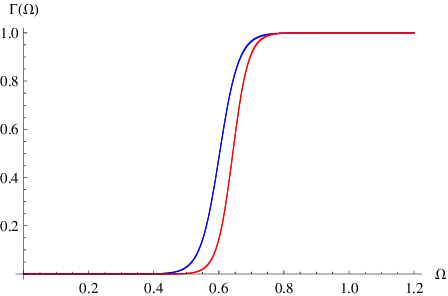}\includegraphics{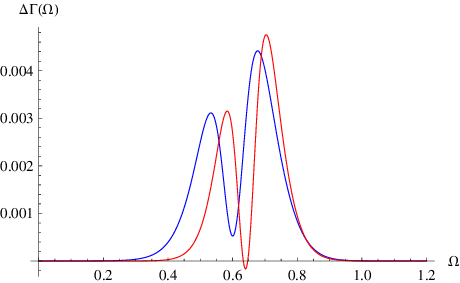}}
\caption{Left panel: Gray-body factors obtained by the 6th order JWKB approach and via the correspondence with QNMs $\ell=3$, $\xi=0.01$ (blue) and $\xi=0.45$ (red); $M=1$. Right panel: Difference between gray-body factors from both methods.}  \label{fig:GBFL3}
\end{figure*}

In the eikonal (high multipole number \( \ell \gg 1 \)) limit, there exists a well-established correspondence between QNMs and the parameters of unstable circular null geodesics~\cite{Cardoso:2008bp}. Specifically, the real part of the quasinormal mode frequency is associated with the angular velocity \( \Omega_c \) of the null geodesic, while the imaginary part is related to the Lyapunov exponent \( \lambda \), which characterizes the instability timescale of the orbit. The leading-order relation takes the form:
\begin{equation}
\omega_{\text{QNM}} \approx \ell\, \Omega_c - i \left(n + \frac{1}{2}\right) \lambda,
\end{equation}
where \( n \) is the overtone number. This approximation provides geometric insight into the origin of quasinormal ringing in the eikonal regime and holds for a wide class of static, spherically symmetric, and asymptotically flat black hole spacetimes. Nevertheless, several exceptions to this correspondence have been identified in the literature~\cite{Konoplya:2017wot,Konoplya:2022gjp,Bolokhov:2023dxq,Konoplya:2025afm}, often arising from cases where the JWKB approximation—on which the correspondence relies—is no longer valid. In particular, this occurs when the centrifugal term dominating the eikonal regime of the effective potential deviates from the standard form $g_{tt}=\ell(\ell+1)r^{-2}$, as happens in various theories with higher-curvature corrections \cite{Konoplya:2020bxa}. Another situation arises when the Taylor expansion near the potential peak fails to match accurately with the WKB series at the boundaries \cite{Bolokhov:2023dxq}.

As a result, the applicability of the correspondence must be verified on a case-by-case basis. In the present context, however, one can easily confirm that the correspondence remains valid even in the presence of quantum corrections, as demonstrated by the analytic expressions derived above.

Finally, by evolving the perturbations to sufficiently late times, as illustrated in Fig.~\ref{fig:TD2}, we observe that the ringdown phase transitions into a power-law decay governed by the relation
\begin{equation}
|\Psi| \sim t^{-(2\ell + 3)}.
\end{equation}
This asymptotic behavior matches the late-time tail first identified by Richard Price for Schwarzschild black holes~\cite{Price:1971fb,Price:1972pw}. Therefore, the decay law appears to exhibit a universal character, remaining unaffected by the quantum correction parameter \( \xi \). Moreover, the time-domain profiles show no signs of growing modes, confirming the stability of the system — consistent with the positive definiteness of the effective potential.

\section{Gray-body factors} \label{sec:GBF}

Gray-body factors exhibit greater stability under small deformations of the background geometry compared to QNMs \cite{Oshita:2023cjz,Oshita:2024fzf,Wu:2024ldo,Rosato:2024arw}. While QNMs—particularly the overtones—can be highly sensitive to localized modifications of the metric, whether near the event horizon (e.g., due to quantum effects) \cite{Konoplya:2022pbc} or in the far zone (e.g., from astrophysical environments) \cite{Barausse:2014tra}, gray-body factors typically remain largely unaffected by such perturbations. This robustness makes them a reliable probe of the global spacetime structure. Furthermore, a correspondence between QNMs and gray-body factors has been established \cite{Konoplya:2024lir}, enabling the latter to be inferred semi-analytically from the former. The correspondence links the fundamental mode $\omega_0$ with gray-body factors $\Gamma_\ell(\Omega)$  via the relation \cite{Konoplya:2024lir}
\[
\Gamma_\ell(\Omega)  = \left(1 + \exp\left[\frac{2\pi\left(\Omega^2 - \text{Re}(\omega_0)^2\right)}{4\, \text{Re}(\omega_0)\, |\text{Im}(\omega_0)|} \right]\right)^{-1} + \mathcal{O}(\ell^{-1}).
\]
This correspondence has been tested in recent literature across a range of models, including those with asymptotically de Sitter geometry and quantum corrections~\cite{Malik:2024cgb,Bolokhov:2024otn,Konoplya:2024vuj,Pedrotti:2025idg,Hamil:2025cms,Lutfuoglu:2025ljm,Dubinsky:2024vbn,Skvortsova:2024msa,Skvortsova:2024msa,Tang:2025mkk}.

As shown in Figs.~\ref{fig:GBFL2} and \ref{fig:GBFL3}, the gray-body factors decrease as the quantum parameter \( \xi \) increases. This behavior can be clearly understood by examining the effective potentials (see Figs.~\ref{fig:gravpot1} and \ref{fig:gravpot2}), which become higher for larger values of \( \xi \). A taller potential barrier allows less radiation to tunnel through, thereby reducing the transmission probability. The gray-body factors calculated using the correspondence with QNMs differ from those obtained via the higher-order JWKB method by only a few percent for \( \ell = 2 \), and by less than one percent for higher multipoles. This indicates that the correspondence provides a reasonably accurate approximation in this context.

\section{Conclusions} \label{sec:Conc}

QNMs of various quantum-corrected black holes—obtained through the inclusion of additional matter fields, higher-curvature corrections, or other effective approaches — have been actively studied in recent years~\cite{Konoplya:2025hgp,Chen:2023wkq,Baruah:2023rhd,Konoplya:2023ahd,Fu:2023drp,Konoplya:2017lhs,Liu:2020ola,Cruz:2020emz,Cuyubamba:2016cug,Konoplya:2017ymp,Xing:2022emg,Daghigh:2020fmw,Piedra:2009pf,Zahid:2024hyy,Zahid:2024hwi,Okyay:2021nnh,Gogoi:2023kjt}. A particularly interesting model was recently proposed in~\cite{Bonanno:2023rzk}, where quantum corrections were incorporated using tools from the renormalization group and gravitational collapse, resulting in a fully regular black hole geometry.

While the QNMs for test scalar, Dirac, and Maxwell fields in this background have been computed in~\cite{Stashko:2024wuq}, our work extends this analysis to include axial gravitational perturbations. We have shown the following:
\begin{itemize}
\item The quantum-corrected black hole behaves as a more efficient oscillator than its classical counterpart; the damping rate is more strongly influenced by quantum corrections than the real part of the frequency.
\item An analytic expression was derived in the eikonal limit, confirming the correspondence between null geodesics and eikonal QNMs.
\item The asymptotic power-law tails of the perturbations are indistinguishable from those of the Schwarzschild black hole.
\item Gray-body factors are suppressed by the quantum parameter \( \xi \), and we demonstrate that they can be accurately estimated using the correspondence with QNMs, as proposed in \cite{Konoplya:2024lir}.
\end{itemize}

Finally, we remark on the expected scale of the spectral modifications induced by the quantum parameter~$\xi$. Within the explored parameter range, the damping rates exhibit deviations of up to $20\%$, while the real parts of the frequencies shift by about $7\%$ relative to their Schwarzschild counterparts. These differences are well above the numerical uncertainties and clearly quantify the impact of quantum corrections on the spectrum.

Beyond their purely theoretical interest, these results may also be relevant in the context of primordial black holes and Planck-scale remnants. Quantum corrections become dominant for masses close to the Planck scale, and the modified grey-body factors may influence the Hawking radiation evaporation of primordial black holes at their late stage.

This work may be further extended by analyzing higher overtones, for which a more precise method such as Leaver's continued fraction technique~\cite{Leaver:1985ax} would be required.

\begin{acknowledgments}
The author would like to thank the anonymous referee for their constructive comments and valuable suggestions, which helped to improve the clarity and quality of this work. The author is grateful to Excellence Project PrF UHK 2205/2025-2026.
\end{acknowledgments}

\bibliography{bibliography}
\end{document}